# Physical foundations of biological complexity

Yuri I. Wolf[1], Mikhail I. Katsnelson[2], Eugene V. Koonin[1,*]

[1]National Center for Biotechnology Information, National Library of Medicine, Bethesda, MD 20894

[2]Radboud University, Institute for Molecules and Materials, Nijmegen, 6525AJ, Netherlands

*For correspondence: koonin@ncbi.nlm.nih.gov

**Abstract**

Biological systems reach hierarchical complexity that has no counterpart outside the realm of biology. Undoubtedly, biological entities obey the fundamental physical laws. Can today's physics provide an explanatory framework for understanding the evolution of biological complexity? We argue here that the physical foundation for understanding the origin and evolution of complexity can be envisaged at the interface between the theory of frustrated states resulting in pattern formation in glass-like media and the theory of self-organized criticality (SOC). On the one hand, SOC has been shown to emerge in spin glass systems of high dimensionality. On the other hand, SOC is often viewed as the most appropriate physical description of evolutionary transitions in biology. We unify these two faces of SOC by showing that emergence of complex features in biological evolution typically if not always is triggered by frustration that is caused by competing interactions at different organizational levels. Competing interactions and frustrated states permeate biology at all organizational levels and are tightly linked to the ubiquitous competition for limiting resources. This perspective extends from the comparatively simple phenomena occurring in glasses to large-scale events of biological evolution, such as major evolutionary transitions. We therefore submit that frustration caused by competing interactions in multidimensional systems is the general driving force behind the emergence of complexity, within and beyond the domain of biology.

**Significance**

Living organisms are characterized by a degree of hierarchical complexity that appears to be inaccessible to even the most complex inanimate objects. Routes and patterns of the evolution of complexity are poorly understood. We propose a general conceptual framework for emergence of complexity through competing interactions and frustrated states similar to those that yield patterns in stripe glasses and cause self-organized criticality. We show that biological evolution is replete with competing interactions and frustration. The key distinction between biological and non-biological systems seems to be the existence of long-term digital memory in the former.

**Introduction**

"All science is either physics or stamp collection". This oft repeated aphorism attributed to Ernest Rutherford tends to annoy scientists to no end due to the flagrant disregard of non-physical science that it espouses. However, if one looks beyond the purported insult, the phrase seems to have a serious, even fundamental meaning. Indeed, it emphasizes the major distinction between those scientific endeavors that strive to explain reality in terms of universal, microscopically reversible laws and do not concern with particularities of the real-world history, and those that depend on history in their explanatory frameworks, such as (most of) biology or geology. The first class of approaches can be collectively denoted physics (1). There is no corresponding single term for the second type of scientific endeavors but for the sake of discussion, we can call them "historical sciences", to emphasize their intrinsic time-irreversibility and dependency on unique events. In historical sciences, universal laws figure only in the background (although the fundamental laws of physics certainly hold), predictability and reproducibility become problematic, and any generalization is suspect (biologists seem to rather delight in saying that "there are exceptions to everything in biology"). Hence the "stamp collection" moniker, whether or not one views it as derogatory.

Currently, fundamental laws of physics are considered to be local in space and time. Beginning with Newton's *Principia* (2), physicists have held that the fundamental laws are expressed in terms of ordinary differential equations, that is, the whole evolution of a (mechanical) system is uniquely determined by coordinates and velocities of all particles at a given time instant. In classical field theory, the fundamental laws are represented by partial differential equations, that is, they are local not only it time but also in space. Indeed, this is the only type of physical laws that are consistent with general relativity theory which shuns instantaneous interaction at finite distances (3). Then, how is it possible that many physical systems have history (in other words, memory of events past), sometimes, going back for as long as billions of years? What are the physical mechanisms that could be responsible for the long-term memory in such systems? Coming back to Rutherford, how does "stamp collection" emerge in the world of physics?

There are major research areas that traditionally fall within the domain of physics, but where history matters. This is the case for all non-ergodic systems of which structural and spin glasses (4, 5) arguably are the best studied class. Glass-like systems are non-Markovian, i.e. are characterized by historical memory whereby the present state of the system depends on the entire previous history, and accordingly, the future states are not precisely predictable. This non-ergodic behavior is caused by competing short and long range interactions which result in frustration and can produce complex patterns (6). Notably, the glass-like phases share with biological systems at least two fundamental features: i) historical memory and the resulting contingency, and ii) complexity. These parallels have been drawn previously by Laughlin, Pines and colleagues who suggested that modern physical theory of glasses might substantially inform different areas of theoretical biology (7, 8). Here we develop the idea of the crucial, central role of competing interactions and frustration in biological evolution by examining specific biological concepts and phenomena, and comparing their behavior to that of glass-like systems. In contrast with the previous analyses (7, 8), which primarily emphasized glass-like features of biological macromolecules (see also (9)), we focus on frustrations and competing interactions in evolutionary dynamics.

Biological evolution undeniably leads to the emergence of complexity (10-16). Moreover, complexity has been proposed as an identifying feature of "artefacts", i.e. systems of (ultimate) biological origin. In a (so far largely gedunken) search of putative extra-terrestrial habitats of life, any objects of "unreasonable complexity" could be the principal signatures of biological activity (17). Complexity is notoriously difficult to define precisely although it seems to be commonly held that "when we see it, we know it". The definitions that appear to be meaningful in biology involve, depending on the level of analysis, the number of evolutionarily conserved nucleotide sites in genomes, number of genes or functional components in organisms or sub-organismal functional systems as well as the hierarchical organization of biological entities, be it functional networks and pathways, cells, organisms or communities (14-18). Perhaps, the most general definition, pathway complexity, driving from the concept of algorithmic complexity in mathematics, includes the number of steps required to create a given object. It has been proposed that entities with a pathway complexity above a certain threshold can only originate from biological processes (17).

The origins of biological complexity have been approached from both biological and physical standpoints. Traditional biological narratives view complex features as adaptations. However, more recently, neutral scenarios for evolution of complexity have been proposed (19-24). Under these models, complexity ensues from evolutionary accumulation of genomic features, such as duplicated genes, introns, or mobile elements. Such features accumulate non-adaptively in organisms with small effective population size, in which purifying selection is too weak to purge them. The genomic embellishments that have not been discarded can gain function, for instance, under the subfunctionalization model of evolution of duplicated genes, when the duplicates differentially lose subfunctions of the ancestral genes. This non-adaptive process translates into "constructive neutral evolution" whereby complexity emerges in the form of interdependence of subfunctionalized components of an evolving cellular system (25-28). However, adaptive origin of complexity cannot be discarded either as suggested, for example, by recent models of prokaryotic genome evolution indicating that acquired genes are, on average, beneficial, conceivably, thanks to the increased functional versatility of more complex microbes (29, 30).

The adaptive or neutral biological models both fail to capture one of the key aspects of the emergence of biological complexity, namely, the striking temporal non-uniformity of the appearance of complex features. The apparently abrupt surges of complexity during evolution are embodied in the concept of major and minor evolutionary transitions (31, 32) and, under a different perspective, in the concept of punctuated equilibrium (33, 34). From a physical standpoint, a rise in complexity is linked to the widespread phenomenon of self-organized criticality (SOC) (35-41) . Self-organized criticality is a property of dynamical systems with extended degrees of freedom and pronounced non-linearity whereby the system goes through serial 'avalanches' separated in time by intervals of stability. The analogy between SOC and punctuated equilibrium in biology is obvious, and has been invoked to account for the origin of biological complexity (37, 38, 40, 41) .

So far, the exact conditions leading to SOC have not been identified despite considerable effort. However, a notable connection has been shown to exist between competing interactions and frustration in spin glasses, on the one hand, and SOC, on the other hand (42, 43). Specifically, it has been shown that SOC is an emergent property of spin glasses with a diverging number of neighbors (43).

Here, we explore the biological implications of these concepts and findings, and conclude that complexity in biological systems emerges in strongly connected multicomponent ensembles, from competing interactions that lead to recurring frustrated states and SOC. We trace these phenomena at all levels of biological organization including both minor and major evolutionary transitions. We further submit that frustration is characteristic of an extremely broad class of hierarchical systems, within and outside biology, and accordingly, appears to be the general source of complexity.

**Glasses, patterns, frustrated states and self-organized criticality**

Geometric frustrations in crystallography (6, 44) can be considered the prototype of frustrations in general. Some types of chemical bonds (e.g., Van der Waals and metallic) tend to the closest possible packing of atoms or ions. The local closest packing of hard discs of equal radii in a two-dimensional Euclidean space is provided by the equilateral triangle, and the global closest packing is the triangular lattice build from such triangles. In this case, there is a unique optimal crystal lattice, with no frustrations involved. In the three-dimensional Euclidean space, the optimal local packing is provided by the regular tetrahedron. However, it is impossible to fill the space with the tetrahedra without voids. As a result, there are infinitely many structures corresponding to the same optimal global packing which is less dense than the optimal local packing; such degeneracy is observed also for higher space dimensionalities as well (45).

Another important source of frustrations are competing interactions caused by the coexistence of several types of chemical bonds (in particular, van der Waals and hydrogen, or metallic, covalent, and ionic). Typically, there is no unique optimal structure with the lowest energy; rather, frustrations lead to quasi-degeneracy. Even elemental solids are usually polymorphic, with different phases, such as graphite, diamond and fullerenes in the case of carbon, between which the energy differences are several orders of magnitude smaller than the energy of each phase. All the richness and diversity of the structures of minerals and inorganic solids and, as discussed below, of organic molecules and biopolymers come from these frustrations. In particular, the differences in free energies of DNA molecules with different nucleotide sequences are orders of magnitude smaller than the total energy of covalent bonds in these molecules. This flatness of the free energy landscape of DNA provides for the existence of a combinatorially large number of quasi-degenerate states, i.e. genomes of different organisms (Schroedinger's famous aperiodic crystal (46) can be naturally interpreted in these terms).

Another notable example is provided by the van der Waals heterostructures, i.e. artificial structures made from different two-dimensional materials; graphene on hexagonal boron nitride (hBN) is the simplest and the best studied case (47). Graphene and hBN have the same crystal structure but with slightly different lattice periods (1.8% larger for hBN). To minimize the energy of interlayer van der Waals interactions, expansion of graphene to equalize the lattice constants is favorable but such expansion costs some energy of interactions between carbon atoms in graphene. As a result, a distinct pattern is formed at small enough misalignment angles (48, 49). Similar physics arises and leads to pattern formation when one graphene layer is rotated with respect to another (50). In these cases, patterns originate from the incommensurability of interactions.

Certain competing interactions result in the formation of glassy states as captured in the concept of *self-induced glassiness* (50-53). For example, for magnetic stripe domains, quasi-chaotic patterns result from competition between short-range but strong exchange interaction which tend to maximize magnetization, both locally and globally, and long-range but weak dipole-dipole interaction requiring that the total magnetization of the system is equal to zero.

The appearance of the glassy state in extremely simple physical systems with only two competing interactions, seems to open the way for understanding the origin of long-time memory, i.e. non-ergodic processes in which history matters, from the vantage point of statistical mechanics. As first clearly introduced for spin-glass by Edwards and Anderson (4), modern physics considers glass to be a distinct state of matter that is intermediate between equilibrium and nonequilibrium (54-56). A characteristic property of glasses is aging, or structural relaxation. Suppose we measure a specific property of an equilibrium phase, liquid or solid, e.g. the resistivity of a metal (or liquid metal). "Equilibrium" means that, when the measurement is repeated after a thermal cycle (slow heating and cooling down to the initial temperature), we obtain the same value of the resistivity. In glass, the measured value would slowly change from measurement to measurement. The potential energy relief (or landscape, to use a term with biological connotations) for glass is a function with many (asymptotically, infinitely many) local minima separated by barriers with an extremely broad energy distribution. Each local minimum represents a metastable state. During its thermal evolution, the system slowly moves from one minimum to another. Importantly, the glass state is non-ergodic: there are many configurations which remain localized in a restricted domain of phase space (54-57). A more or less complete formal theory (based on ideas of Parisi (58)) exists only for an artificial, mean-field style model with infinitely long-range, independent, random interactions (Sherrington-Kirkpatrick model). Within this theory, the state of the glass is characterized by an "order parameter" with continuously many components, labeled by a real number $x \in (0,1)$ (58). This number can be represented as an infinite, non-periodic binary fraction, such as 0.10001110…, where 0(1) corresponds to the choice of bifurcation on the complex energy relief when cooling down from the equilibrium liquid state. This process of thermal evolution is naturally described in terms of ultrametricity. This is a coarse grain, "topological" description of the evolution of the system via bifurcations that does not require detailed knowledge of the heights of barriers, rates of transitions and other characteristics (56). This picture is consistent with the principle of SOC: the SOC dynamics includes avalanches of all sizes distributed by a power-law and contains a slow component corresponding to the $1/f$ noise (42). The existence of long-range interactions competing with short-range ones is essential for the emergence of SOC; in systems with short-range interactions only, SOC is not observed (43).

One of the formal criteria of the glass state is "universal flexibility" (59). Because this is the criterion used in the theory of self-induced glassiness (51-53) that is directly relevant for the present analysis, it merits a brief description. Consider a configuration (of spins, atomic positions, dipolar moments or other parameters) that is characterized by a function $\phi(x)$ where $x$ is $d$-dimensional vector characterizing a position in space (in most physical applications, $d = 2$ or 3). The energy of this configuration is given by its Hamiltonian $H[\phi(x)]$ and free energy

$$F = -T \ln \int D\phi \exp(-H[\phi(x)]/T) \qquad (1)$$

where $T$ is the absolute temperature (we put Boltzmann constant equal to one) and $\int D\phi$ represents summation over all possible configurations. Let us add interaction with another configuration $\sigma(x)$:

$$H[\phi(x)] \to H_g[\phi(x)] = H[\phi(x)] + \frac{g}{2}\int dx[\phi(x) - \sigma(x)]^2 \qquad (2)$$

and calculate the free energy $F_g$ replacing $H[\phi(x)] \to H_g[\phi(x)]$ in Eq.(1). Then, let us consider two transitions: thermodynamic limit $V \to \infty$, where $V$ is the volume of the system, and the limit of infinitely weak coupling $g \to +0$. If these limits do not commute, i.e.

$$\lim_{g \to +0} \lim_{V \to \infty} \frac{F_g}{V} \neq \lim_{V \to \infty} \lim_{g \to +0} \frac{F_g}{V} \qquad (3)$$

for a macroscopically large number of configurations $\sigma(x)$, then, the system is glass, i.e. a non-ergodic state with memory. Physically, this means that the energy landscape for the glass represents a "universal mapping function", so that for many $\sigma(x)$, there exists a part of the landscape that is minimized by the choice $\phi(x) = \sigma(x)$. The criterion (3) is fulfilled, under certain conditions, in some *non-random* systems with competing long-range and short-range interactions, such as ferromagnetic thin films (51-53). This criterion appears to be consistent with SOC because, in such systems, the effective connectivity is infinitely large (due to the long range character of dipole-dipole interactions).

The critical dimensionality, above which the Parisi description holds, is 6 (60). In biological evolutionary dynamics, the dimensionality of the configuration space (fitness landscape) is typically very high: for example, numerous genes in a genome and numerous sites in each gene can be considered separate dimensions (61). Therefore, biological evolution can be expected to follow the chain of causation: competing interactions → frustration → glass-like state → non-ergodicity → SOC → evolutionary transitions/"punctuated equilibrium" (Figure 1).

When $\sigma(x)$ is simply one specific function, equation (3) is equivalent to the condition of spontaneously broken symmetry in the Landau theory of second-order phase transitions (62, 63), where $h(x) = g\sigma(x)$ plays the role of external field conjugated to the order parameter $\phi(x)$. Conceivably, for some systems, the criterion (2), (3) can be satisfied neither for an "almost arbitrary" function $\sigma(x)$ as in glasses nor for a single function as in conventional second-order phase transitions, but rather for a sufficiently rich but limited set of functions. Such systems would spontaneously "glue" to selected configurations from some "library" to form a complex but not completely chaotic pattern. For example, in thin ferromagnetic films, the "striped glass" phase is not completely random but rather is characterized by a specific spatial scale; this is the length where strong but short-range and weak but long-range interactions have the same order of magnitude (52). Such patterned glasses are likely to yield better models for biological phenomena than classical glass. Here, we assume that there are many distinct "attractors" and that the energy landscape of the system consists of glass-like parts separated by gaps. This model immediately invokes an analogy with pattern recognition in learning theory that has been successfully studied within the framework of the spin-glass formalism (5, 57). A clear and perhaps fundamental analogy from evolutionary biology is a typical, rugged fitness landscape with elevated areas (peaks and plateaus) of high fitness, where an evolving population can travel either upwards, under the pressure of selection, or horizontally in a (quasi)neutral evolutionary regime, separated by valleys of low fitness that can be crossed only by genetic drift (thermal fluctuation) (61).

"Adaptation is localization in sequence space" (64), i.e. fixation of distinct genotypes that makes evolution along unique trajectories. The feasibility of such fixation depends on the replication fidelity which has to exceed a critical threshold (often called "Eigen threshold", after Manfred Eigen, the originator of the replicator theory (65)) and on the complexity of the fitness landscape. Below the threshold, in a simple, single-peak fitness landscape, a replicator system devolves into a random population of sequences. In physical terms, such a population represents an ergodic system whereas an evolving system with fixation on a rugged fitness landscape is non-ergodic (Figure 2). There is a rich enough set of attractors $\{\sigma(x)\}$ in the sequence space but the standard glass model, where attractors represent a substantial fraction of the configurations so that the landscape is a virtual continuum, does not appear directly relevant for biological evolution. The evolutionary process is made possible by the existence of distinguishable, discrete states. Indeed, genetic information can be changed only in discrete steps, one nucleotide or one codon at a time (at the finest granularity), as opposed to the effective continuity (glass-like character) of the phenotype. This fundamental distinction between the genotype and the phenotype underlies the "central dogma" of molecular biology: the fundamental carriers of biological information (nucleic acids) have to be digital whereas operational parts, such as proteins, are analog devices (66). Discreteness of biological systems requires that any relevant attractor has a *finite* basin such that the basins of different attractors should be separated by sufficiently high barriers. RNA molecules can be viewed as occupying an intermediate position between the discreteness of the DNA state and "glassiness" of proteins, being capable of functioning both as information carriers and as operational devices. Hence the primordial RNA world scenario. Thus, direct analogies between biology and physics of glasses (7, 8) seem to be oversimplified. The essential concept for biological evolution appears to be that of pattern formation as observed in stripe glasses.

**Competing interactions and frustrated states as drivers of biological evolution**

Competing interactions are apparent between all kinds of biological entities, and frustrated states seem to emerge at all biologically relevant levels of organization (Table 1). Arguably, the lowest level of specific biological complexity is folding of nucleic acid and protein molecules into unique, biologically functional three-dimensional structures (67). From the evolutionary perspective, the beginning of life can be most plausibly associated with the appearance of the first RNA molecules (ribozymes) endowed with RNA polymerase activity within the hypothetical, primordial RNA World (68). Not unexpectedly, laboratory experiments that attempt to select for RNA molecules with polymerase activity show that it can be achieved (so far, to a limited extent) only by structurally highly elaborate RNAs (69-71). The competition between short-range and long-range interactions is plainly apparent in protein and RNA molecules, and is the defining factor of folding that underlies all molecular functions. The transition from the primordial RNA World to the modern-type DNA-RNA-protein biology appears analogous to symmetry breaking whereby the "glassy" landscape of catalytic RNAs is partitioned into the patterned, digital genotype (DNA) and the continuous, analog phenotype (proteins and structural RNAs).

Moving up a level, in macromolecular complexes that, in actuality, perform most if not all, biochemical functions in cells and viruses (72, 73), the competition between interactions within individual macromolecules and those between subunits that lead to complex formation is equally obvious. Examples abound, suffice it to point out the conformational changes in both ribosomal RNA and proteins during ribosome subunit formation (74, 75), in transcription factors upon DNA binding (76, 77), and in virion proteins during morphogenesis of virus particles (78).

Furthermore, allosteric regulation of enzymatic complexes involves transition between macromolecular conformations with close free energies but distinct biological properties (67). The existence of many conformations with comparable energies in biological complexes has the same cause as the polymorphism of simple (including elemental) inorganic solids mentioned above, namely, competing interactions.

On another plane of biological organization that is unique to living matter, competing interactions can be conceptualized as selection pressures that act in opposite directions. A complex fitness landscape with many basins of attraction (that is, with many evolutionary strategies with more or less the same adaptive efficiency) can result only from an interplay of these competing factors. In particular, such conflicting selective processes are a key, inherent component of host-parasite coevolution. Emergence of genetic parasites appears to be inevitable in all evolving replicator systems because it can be shown that parasite-protected systems are evolutionarily unstable (79, 80). Moreover, genetic parasites with different reproduction strategies (viruses, plasmids, transposons and more) accompany (nearly) all cellular life forms (81-84). The frustrated state of a host-parasite system is caused by a complex interplay between the parasite replication, the host replication and the interactions that stabilize the host-parasite system as a whole. The conflicts between the selective factors that operate on each of these levels appear to be a major, perhaps the principal driver of evolution of these systems (85). Computer simulations under a wide range of conditions show that, in a well-mixed replicator system, parasites overwhelm the hosts, resulting in the eventual collapse of the entire host-parasite system (80, 86-89). In contrast, compartmentalization (that, in specific terms, could represent, for instance, partitioning of replicator ensembles between different microcompartments in an inorganic compartment network or simply separation in a viscous medium) stabilizes the system and leads to diversification and evolution of complexity (86-88). Effectively, the outcome of host-parasite coevolution in such modeling studies is pattern formation, a typical consequence of frustration in glass-like states.

Compartmentalization is arguably the simplest, most fundamental effect of host-parasite conflicts but, in all cellular life forms, these conflicts also drive the evolution of versatile host defense systems and counter-defense systems in parasites, another prominent and ubiquitous manifestation of biological complexity (90-93). In the course of evolution, conflicts between hosts and parasites are resolved into multiple, distinct, stable evolutionary regimes. These regimes span the entire range of host-parasite relationships, from highly aggressive parasites, such as lytic viruses, that kill the host and move to the next one, to cooperative elements, such as many plasmids, that often provide beneficial functionalities to the host (83, 84). This persistent diversity of host-parasite interactions is a major part of biological complexity at the level of ecosystems and the entire biosphere.

Even more generally, it seems to be no exaggeration to state that frustration caused by intergenomic conflicts drives the evolution of all biological complexity (85, 94). The genomes of all cellular life forms contain multiple inserted MGE genomes which in multicellular eukaryotes (animals and plants) account for the majority of the genome sequence (95-97). The competing interactions causing frustration are particularly obvious in the case of MGE with dual activities, such as toxin-antitoxin (TA) and restriction-modification (RM) modules in prokaryotes (98-101). On the one hand, TA and RM systems protect the hosts from other, more aggressive parasites, primarily, viruses, but on the other hand, they themselves behave as MGE. Frustration in bacteria and archaea that harbor TA and RM (that is, nearly all bacteria and archaea) is manifest at several levels. The TA and RM systems compete, on the one hand, with viruses, which they attack and hence protect the host, and on the other hand, with the host itself, which they kill when it "attempts" to get rid of these elements. Another part of this gamut are plasmids on which RM and TA modules are typically transferred. The outcome of these complex networks of competition is the stabilization of the entire host-parasite system in which components with all types of reproduction strategies persist indefinitely. In other words, host-parasite coexistence translates into persistence of biological complexity at the ecosystem level. A striking feature of the competition networks is the "guns for hire" phenomenon, i.e. shuttling of the same active components, such as nucleases involved in transposition, between defense systems and MGE (102, 103).

**Evolutionary transitions and major innovations driven by competing interactions**

Competing interactions and frustration naturally apply to evolutionary transitions and, more generally, major evolutionary innovations (MEI). The concept of major transitions in evolution (MTE) developed by Maynard Smith and Szathmary defines a distinct class of MEI that involve evolutionary transitions in individuality (ETI) (31, 32). A classic example of a MTE is the origin of multicellular organisms from unicellular life forms but MTE, although not numerous, punctuate the entire history of life (Table 1). The key tenet of the MTE concept is that, within its framework, the transitions are construed not simply as MEI but rather meet strict criteria that make them akin to phase transitions in physics. Thus, every MTE is a MEI but not every MEI is an MTE. The signature feature of MTE is ETI, which involves a change in the level of selection, e.g. from a single cell to an ensemble of cell (a multicellular organism) (Table 1 and Figure 1). The second signature of MTE, *senso* Szathmary, is that each transition is associated with the emergence a new type of information storage, use and transmission (e.g. multicellularity is linked to the rise of epigenetic informational systems) (32). Competing interactions and/or levels of selections are immediately apparent in each MTE (Table 1).

Starting from the most obvious, evolution of multicellularity involves the intrinsic conflict between selection forces acting at the level of individual cells and those that are manifest at the level of cellular ensembles or tissues. Clearly, to maintain the functionality of a multicellular organism, the proliferation of individual cells has to be tightly controlled.

Moving back in time, we know little about the origin of the first cells. It is nevertheless difficult to imagine an evolutionary scenario in which the emergence of cells was not preceded by an evolutionary stage at which all genetic information was encoded in small elements that resembled modern MGE (104-106). Subsequent evolution involved accretion of such elements to form large genomes such as those of modern prokaryotes. Under this scenario, the emergence of cells involved competition between the selection factors that affects individual genetic elements and those that act on ensembles of such elements that formed cellular genomes. Once again, replication of individual elements has to be subdued for the ensemble – the cell - to function. A closely similar evolutionary scenario has been implemented in a recent mathematical model of primordial cell evolution (107).

The next MTE, the origin of eukaryotes, is associated with endosymbiosis that gave rise to the mitochondria and hence involved the inevitable conflict between the endosymbiont and the evolving eukaryotic cell that required coordinated reproduction of the host and the symbiont to survive and function (108-110). The same conflict is inherent in the evolution of photosynthesizing algae via the cyanobacterial endosymbiosis that gave rise to the chloroplast. The frustration caused by host-symbiont conflicts in these MTE was resolved by the formation of the stable symbiotic associations, but the conflicts linger, e.g. in the form of mitochondrial diseases (111)and frequent lysis of mitochondria which in some organisms results in frequent insertion of non-functional mitochondrial DNA into the host genome (112).

The later MTE that led to the emergence of eusocial animals and societies also clearly involve competition between individuals and collective, or between collectives at different levels of organization. Generally, it appears that for ETI, which is the defining feature of MTEs, competing interactions between entities at different organizational levels are the intrinsic driving factor.

Notably, the conflicts between the different levels of individuality in MTE appear to be intertwined with host-parasite conflicts. Indeed, mathematical models of the evolution of multicellularity suggest that defense against viruses could be a major driving factor of this MTE. In particular, parasite pressure drives the evolution of programmed cell death, a distinct form of defense that functions only in the presence of cell aggregation and appears to promote emergence of multicellularity (113).

Genetic parasites appear to have played important roles also in earlier MTE, in particular, the origin of eukaryotes that is thought to have been accompanied by massive invasion of introns from the endosymbiont into the host genome (109, 110, 114). This explosive invasion of parasitic elements into the genomes of the emerging eukaryotes could have triggered the evolution of several key features of eukaryotic cells that are central to the dramatic increase in cellular organization compared to the prokaryotic ancestors, including the nucleus, the spliceosome and the ubiquitin signaling network (115). However meager our knowledge of the origins of the first cells might be, it appears all but certain that conflicts between selfish and "cooperative" genetic elements played an important role. Thus, all MTE seem to involve more than one type of competing interactions.

Every MTE is a MEI but not the other way around. Nevertheless, it can be argued that MTE differ from other MEI in degree rather than in kind. The MEI that are not associated with ETI nevertheless involve local transitions in individuality, i.e. emergence of new, complex functions through evolutionary fixation of new interactions between genes, such as photosystems components in the case of photosynthesis (116-118) or enzymes of methanogenic pathways in the case of archaeal methanogenesis (119, 120). It can be argued that any MEI involves emergence of a new unit of selection (Darwinian individual) if not a new level of selection (class of Darwinian individuals). Accordingly, it seems that competing interactions and frustration are inherent in all MEI.

**Cancer, aging and death**

The conflict between the propagation of individual cells and the maintenance of stable tissues and organs in multicellular organisms is resolved via multilayer systems of controls of cell proliferation, another striking manifestation of biological complexity. However, an alternative and common resolution of this conflict triggered by impairment of such mechanisms is unchecked proliferation of cheater cells that can lead to tumorigenesis and, in particular, cancer in animals (121). The emergence of cheaters in cell ensembles appears inevitable for the same reasons that make the emergence of parasites intrinsic to any replicator system.

Furthermore, aging, apparently an inherent feature of multicellular organisms, seems to be caused by the same conflict (122). This is the case because elimination of senescent cells that have accumulated deleterious somatic mutations via competition with high-fitness cells conflicts with the propensity of the latter to divide uncontrollably and thus form tumors. In other words, there is an inherent conflict between the fitness ("vigor") of an individual cell that consists in high division rate and cellular cooperation that requires limiting that rate or eliminating division altogether.

**Frustration as the key factor underlying all complexity in nature and the specifics of biological evolution**

Although unifying explanations of universal phenomena are inherently dangerous, it does appear plausible that complexity can emerge only in non-ergodic systems, and non-ergodicity is caused by competing interactions. Furthermore, the competition between short-range and long-range interactions can, under additional constraints, result in SOC. This general perspective on the origin of complexity seems to explain all types of patterns existing in nature, from stripe glasses to planetary systems and galaxies (Figure 1). However, contrary to some general statements (41), there is more to the evolution of complexity than SOC . Importantly, SOC does not lead to hierarchical complexity: fractal patterns produced by SOC are not genuinely complex because, by definition, they appear the same at all spatial and/or temporal scales. In biology, the patterns observed on different levels of organization, such as macromolecules, cells, multicellular organisms, populations and ecosystems, are different.. These levels of organization are linked through MTE which, as argued above, are driven by competing interactions and the resulting frustration. Indeed, the main message of the present work is that this appears to be the universal mechanism driving the evolution of hierarchy in nature. Here, we do not present the actual mathematical theory of frustration-driven evolution. It seems likely that physical and mathematical ideas and formalisms beyond those currently known are necessary to develop such a theory (123).

What are the specific, defining features of life? The distinction of biological systems is certainly not that "nothing in biology makes sense except in the light of evolution" (124), whereas in the rest of the world, "everything does". As discussed above, outside of the quantum realm, the world is full of non-ergodic systems, and conceivably, the universe itself can be properly described only in the light of its evolution over the 13.8 billion years since the Big Bang (125). Yet, the level of complexity and elaboration characteristic of living things is unmatched by anything outside biology. The key difference between biological entities and inanimate non-ergodic, complex systems seems to be the unique biological memory mechanism. Specifically, this mechanism involves replication of digital information carriers (nucleic acids) that memorize the patterns emerging from competing interactions at different levels and propagate that memory with sufficient fidelity to allow selection. Attempts to define life at a fundamental level could be viewed as philosophical exercises of limited interest (126-128). Nevertheless, it does appear that complexity emerging from competing interactions, combined with memory perpetuated by replication of information carriers, underlies all life, and conversely, any system endowed with these properties will qualify as living.

55. Das SP (2004) Mode-coupling theory and the glass transition in supercooled liquids. *Rev. Mod. Phys. 76, 785-851.* 76:785-851
56. Rammal R, Toulouse G, & Virasoro MA (1986) Ultrametricity for physicists. *Rev. Mod. Phys.* 58:765-788
57. Mezard M, Parisi G, & Virasoro MA eds (1987) *Spin Glass Theory and Beyond* (World Scientific, Singapore).
58. Parisi G (1980) A sequence of approximated solutions to the S-K model for spin glasses. *J. Phys. A* 13:1101-1112
59. Monasson R (1995) Structural glass transition and the entropy of the metastable states. *Phys. Rev. Lett.* 75: 2847-2850
60. Wang W, Moore MA, & Katzgraber HG (2017) Fractal dimension of interfaces in Edwards-Anderson spin glasses for up to six space dimensions. *arXiv:1712.04971*
61. Gavrilets S (2004) *Fitness Landscapes and the Origin of Species* (Princeton University Press, Princeton).
62. Landau LD & Lifshitz EM (1980) *Statistical Physics* (Pergamon, Oxford).
63. Ruelle D (1999) *Statistical Mechanics: Rigorous Results* (World Scientific, Singapore).
64. Nowak MA (2006) *Evolutionary Dynamics: Exploring the Equations of Life* (Belknap Press, Cambridge, MA).
65. Koonin EV (2015) Why the Central Dogma: on the nature of the great biological exclusion principle. *Biol Direct* 10:52
66. Finkelstein AV & Ptitsyn OB (2002) *Protein Physics: A Course of Lectures* (Academic Press, New York).
67. Atkins JF, Gesteland RF, & Cech TR eds (2010) *RNA Worlds: From Life's Origins to Diversity in Gene Regulation* (Cold Spring Harbor Laboratory Press, Cold Spring Harbor, NY).
68. McGinness KE & Joyce GF (2003) In search of an RNA replicase ribozyme. *Chem Biol* 10(1):5-14
69. Horning DP & Joyce GF (2016) Amplification of RNA by an RNA polymerase ribozyme. *Proc Natl Acad Sci U S A* 113(35):9786-9791
70. Samanta B & Joyce GF (2017) A reverse transcriptase ribozyme. *Elife* 6
71. Gavin AC*, et al.* (2002) Functional organization of the yeast proteome by systematic analysis of protein complexes. *Nature* 415(6868):141-147
72. Krause R, von Mering C, & Bork P (2003) A comprehensive set of protein complexes in yeast: mining large scale protein-protein interaction screens. *Bioinformatics* 19(15):1901-1908
73. Shajani Z, Sykes MT, & Williamson JR (2011) Assembly of bacterial ribosomes. *Annu Rev Biochem* 80:501-526
74. Kim H*, et al.* (2014) Protein-guided RNA dynamics during early ribosome assembly. *Nature* 506(7488):334-338
75. Wolberger C (1999) Multiprotein-DNA complexes in transcriptional regulation. *Annu Rev Biophys Biomol Struct* 28:29-56
76. Ogata K, Sato K, & Tahirov TH (2003) Eukaryotic transcriptional regulatory complexes: cooperativity from near and afar. *Curr Opin Struct Biol* 13(1):40-48
77. Perlmutter JD & Hagan MF (2015) Mechanisms of virus assembly. *Annu Rev Phys Chem* 66:217-239
78. Koonin EV, Wolf YI, & Katsnelson MI (2017) Inevitability of the emergence and persistence of genetic parasites caused by thermodynamic instability of parasite-free states. *Biol Direct* 12:31 (
79. Szathmary E (2000) The evolution of replicators. *Philos Trans R Soc Lond B Biol Sci* 355(1403):1669-1676
80. Holmes EC (2009) *The Evolution and Emergence of RNA Viruses* (Oxford University Press, Oxford).
20

**Figure legends**

Figure 1. Competing interactions and frustrated states as the ultimate driver of the evolution of complexity.

Figure 2. Ergodic and non-ergodic evolutionary dynamics

Table 1

**Competing interactions and frustrated states in biological evolution**

| System | Frustration-producing factors (competing interactions) | Emergent functional and evolutionary features |
|---|---|---|
| RNA | Short range (within stem local hydrogen bonding, stacking) vs long range (long-distance hydrogen bonding, salt bridges) interactions between nucleotides. | Complex 3-dimensional structures including ribozymes. |
| Proteins | Short range (Van der Waals) vs long range (hydrogen bonds, salt bridges) interactions between amino acid side chains.. | Stable conformations and semi-regular patterns in protein structures. Allostery enabled by transitions between energetically quasi-degenerate conformations. |
| Macromolecular complexes | Within subunit vs between subunit interactions | Elaborate complex organization, in particular, nucleoproteins (ribosomes, chromatin) |
| **Cells[a]** | **Membranes (confinement of chemicals) vs channels/pores (transport of chemicals)** | **Compartments and cellular machinery dependent on electrochemical gradients** |
| **Autonomous (hosts) and semi-autonomous (parasites) replicators** | **Replicator vs parasite genomes** | **Self vs non-self discrimination and defense; complex genomes of increasing size; primitive cells** |
| **Autonomous (hosts) and semi-autonomous (parasites) reproducers/replicators** | **Host cells and viruses** | **Infection mechanisms, defense and counter-defense systems, evolutionary arms race; contribution to the origin of multicellular life forms.** |
| Autonomous (hosts) and semi-autonomous (parasites) reproducers/replicators | Host cells vs transposons | Intra-genomic DNA replication control; evolutionary innovation through recruitment of transposon sequences |
| Autonomous (hosts) and semi-autonomous (parasites) reproducers/replicators | Host cells vs plasmids | Beneficial cargo genes, plasmid addiction systems, efficient gene exchange and transfer mechanisms |
| **Emerging eukaryotic cells** | **Host (archaeal) cells vs endosymbiont (a-proteobacteria, proto-mitochndria)** | **Eukaryotic cells with complex intracellular organization** |
| **Communities of unicellular organisms** | **Individual cells vs cellular ensembles** | **Information exchange and quorum sensing mechanisms; replication control, programmed cell death, multicellularity** |
| Multicellular organisms | Soma vs germline | Complex bodies, tissues and organ differentiation, sexual reproduction |
| Multicellular organisms | Dividing vs quiescent cells | Aging, cancer, death |
| **Populations** | **Individual members vs groups** | **Population-level cooperation; kin selection; eusocilaity** |
| Populations | Males vs females (partners with unequal parental investment) | Sexual selection, sexual dimorphism |
| Biosphere | Species in different niches | Interspecies competition, host-parasite and predator-prey relationships, mutualism, symbiosis |
| **Societies**[b] | … | … |

[a]Those competing interactions and frustrated states that are deemed to directly contribute to MTE are shown in bold

[b]We refrain from specifying the conflicts that drive the origin and evolution of human societies

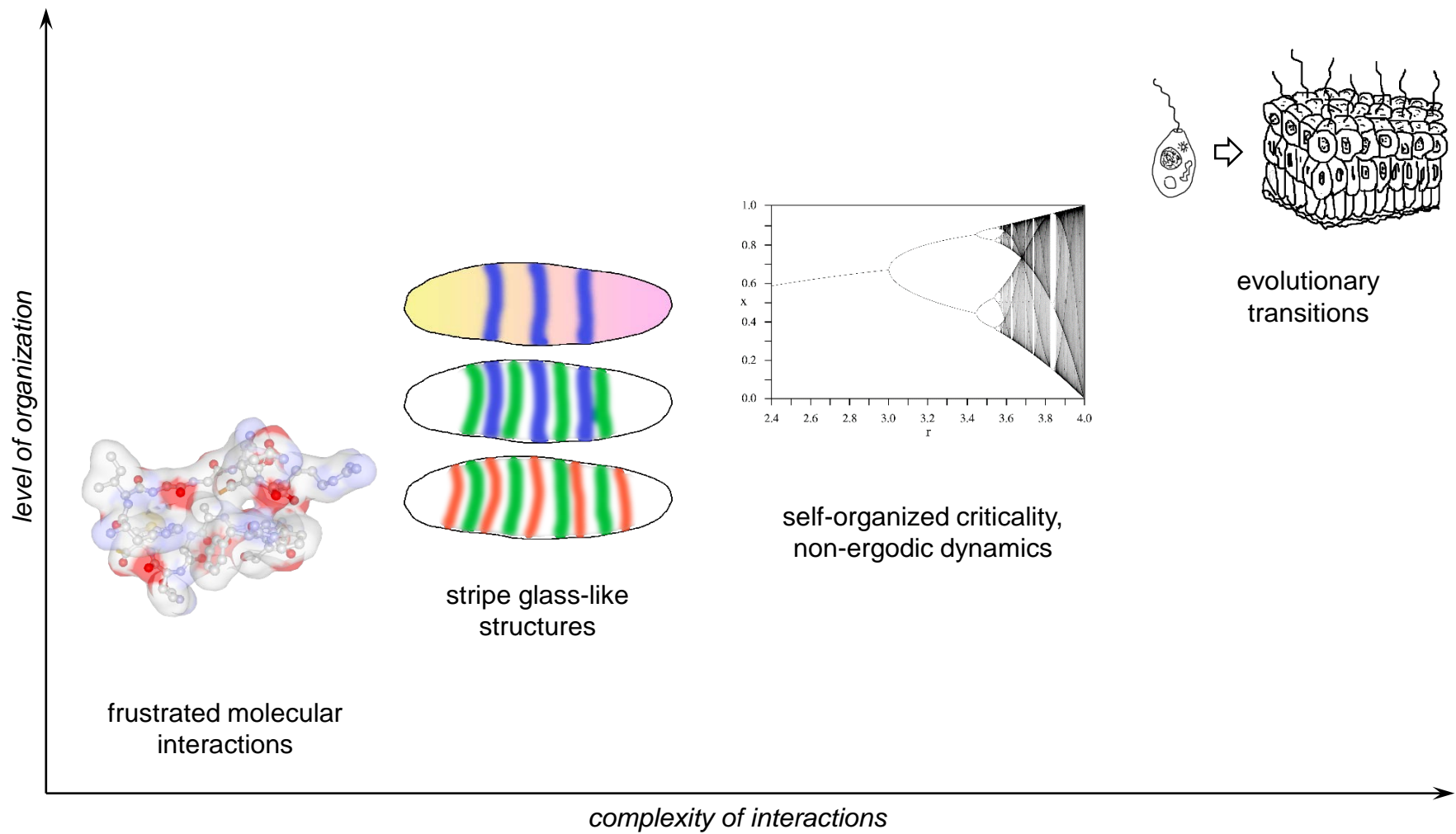

Figure 1

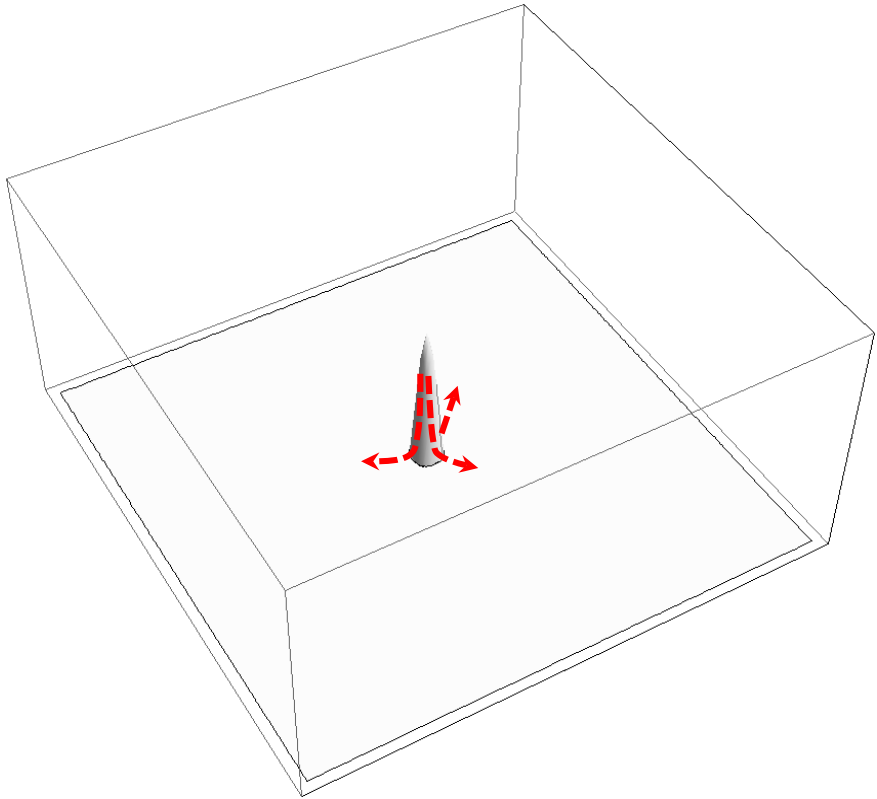 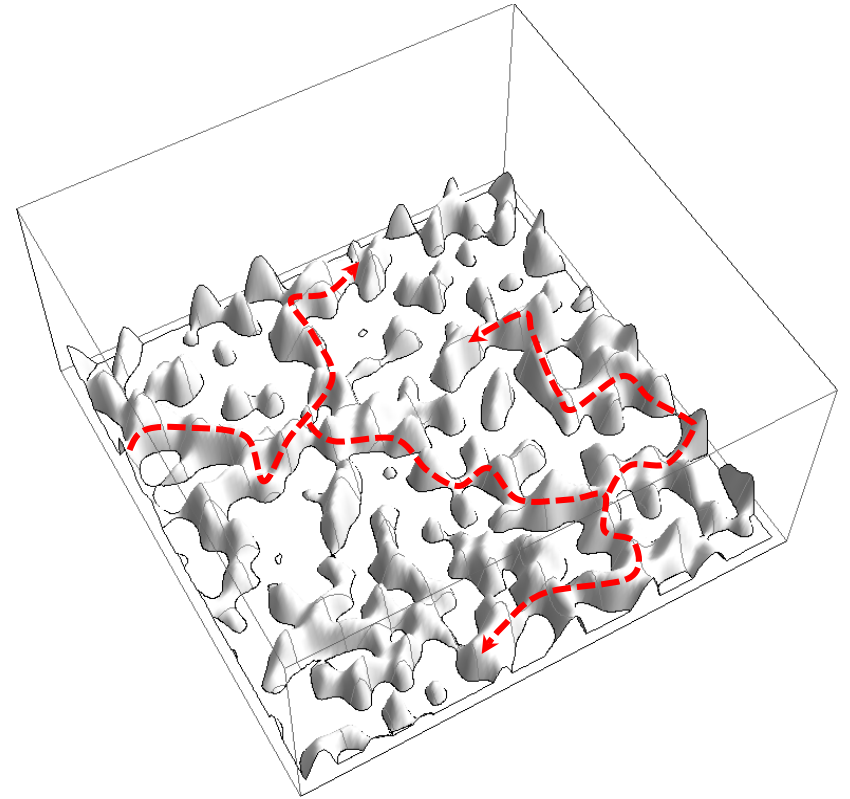

Eigen's single-peak fitness landscape with predictable evolution (survival of the master copy below the error catastrophe threshold or extinction above the threshold)

Complex critical percolation fitness landscape with non-ergodic (divergent) evolution

Figure 2